# Experimental determination of emission cross sections for electron induced processes in a supersonic argon jet

Yu. S. Doronin, A.A. Tkachenko, V. L. Vakula, G. V. Kamarchuk

B. Verkin Institute for Low-Temperature Physics and Engineering

of the National Academy of Sciences of Ukraine,

Kharkiv 61103, Ukraine

E-mail: doronin@ilt.kharkov.ua

## Abstract

The article discusses an original spectroscopic method for the study of supersonic argon jets in atomic and cluster flow regimes. The method is based on absolute measurements of the VUV radiation flux from a supersonic argon jet induced by a 1 keV electron beam. The integral radiation flux and spectral distribution of the radiation flux density of a supersonic argon jet in the range of 50-150 nanometers were measured, which made it possible to determine the density of the non-condensed atomic component, the fraction of condensate, and the density of clusters in a supersonic argon jet at given parameters of its flow into vacuum. The obtained parameters also allowed us to estimate the emission cross sections for the argon resonance lines ArI ($\lambda$=104.8 nm, $\lambda$=106.7 nm) and the cluster continuum $\lambda$=127 nm.

1. ## Introduction

Supersonic flows of inert gases have been actively used in many fields of science and technology for several decades [1-4]. To fully exploit the potential of supersonic flows in science and engineering, it is necessary to understand and control their parameters, such as atom concentration and condensate fraction, cluster density and size, and their structure and size distribution. It is also important to know how these parameters evolve with changes in the conditions of supersonic jet flow in a vacuum. Studying the physical and chemical properties of nanoclusters covering a wide range of particle sizes, from dimers to nanocrystals, is a fascinating and challenging task. Solving this problem requires the use of complex equipment and the latest experimental methods, as well as the development of theoretical models of jet flow with condensation. [5].

In recent years, exciting works have been published in which the authors determined some of the above parameters of supersonic jets by combining several methods of optical diagnostics. For example, Rayleigh scattering, based on elastic scattering of light by particles, provides information on the size and density of gas clusters [6]. Mie scattering provides an estimate of the size distribution of clusters [7], and laser-induced fluorescence (LIF) helps to measure the density and flow pattern of a supersonic gas jet [8]. These methods use rather complex, highly sensitive equipment and require time-consuming analytical modelling and mathematical calculations.

## 2. Experiment

An electron-excited supersonic jet of argon gas is a compact, intense source of vacuum ultraviolet radiation. The interaction of incident electrons with argon atoms and clusters in the jet leads to their excitation and ionisation. The subsequent processes involving atomic and molecular ions, atoms in ground and excited states, and excited and ionised clusters can be studied by luminescence spectroscopy. In the atomic mode of a supersonic argon jet, VUV radiation is mainly generated by resonance transitions between the ground and excited states of atoms and ions. In the cluster mode, as the number and size of clusters in the jet increase, the total VUV radiation flux is dominated by continua emitted by neutral and charged excimer complexes formed in the clusters.

Figure 1 shows the emission spectrum of a supersonic argon jet in the wavelength range 90-150 nm at gas pressure at the nozzle inlet $P_0 = 0.1$ MPa and temperature $T_0 = 300$ K. The supersonic argon jet was crossed at a right angle by an electron beam at a distance L = 30 mm from the nozzle outlet, where the clustering process was practically complete. The contribution of secondary processes to the intensity of the emissions under study was significantly suppressed. The electron energy was 1 keV, and the electron beam current was 20 mA. The spectrum contains resonance lines emitted by argon ions Ar II ($\lambda$ =92.0 nm, $\lambda$=93.2 nm) and atoms Ar I ($\lambda$=104.8 nm, $\lambda$=106.7 nm), as well as cluster continuum with maxima $\lambda$=109 nm (W) emitted by excimers from partially vibrationally relaxed states and continua emitted by neutral $(Ar_2)^*$ $\lambda$=127 nm and charged excimer complexes $(Ar_4^+)^*$ $\lambda$=137 nm in the vibrationally relaxed state [9]. It should be noted that the continuum $\lambda$=127 nm is present in the emission spectra of most plasma VUV sources

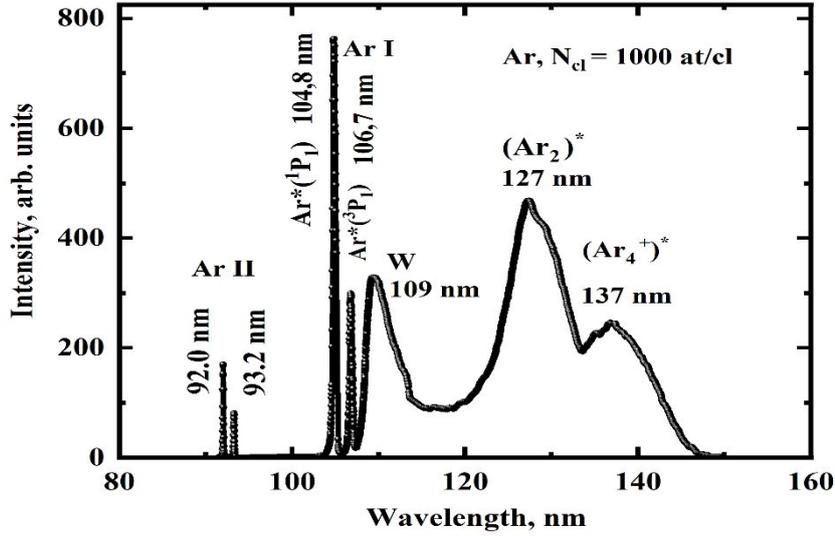

Fig. 1. Spectrum of radiation from a supersonic argon jet, electron energy 1 keV, electron beam current 20 mA, gas pressure and temperature at the nozzle inlet $P_0 = 0.1$ MPa and $T_0 = 180$ K

operating at sufficiently high argon pressures [10] and in the luminescence spectra of liquid and solid argon [11, 12]. At pressure $P_0 = 0.1$ MPa and argon temperature at the nozzle inlet $T_0 > 400$ K, there are no cluster emissions in the spectrum and only Ar II and Ar I lines are registered, which indicates the atomic composition of the supersonic argon jet.

By a specially developed method, we measured the integral radiation flux in absolute units. We determined the distribution of the radiation flux density from the electron-excited supersonic argon jet across the entire wavelength range under investigation [13]. The integral radiation flux Q (photon/s·sr) of a supersonic argon jet at given parameters of gas flow through the nozzle is measured in absolute units in the wavelength range 50-200 nm by a silicon detector SXUV-100 with known spectral sensitivity. Simultaneously, the relative intensity distribution $I(\lambda)$ in the argon jet emission spectrum is recorded by a vacuum monochromator in the same spectral range. The spectrum obtained was corrected, taking into account the efficiency of the vacuum monochromator and the background signal. After determining Q and $I(\lambda)$, the spectral distribution of the radiation flux density $Q(\lambda)$ of the supersonic argon jet in the whole investigated wavelength range is found:

$$Q(\lambda) = \frac{Q \cdot I(\lambda)}{\int_{50}^{200} I(\lambda) d\lambda}, \#(1)$$

where $\int_{50}^{200} I(\lambda)d\lambda$ is the area occupied by the spectrum in the 50-200 nm region.

This method allowed us to measure the absolute intensities of the spectral components present in the spectrum in Fig. 1 as the parameters of the supersonic argon jet flow changed. Fig. 2 shows the dependence of the radiation flux of the Ar II line 92 nm on the gas temperature at the nozzle inlet. The supersonic argon jet crossed an electron beam at a distance of 30 mm from the nozzle outlet.

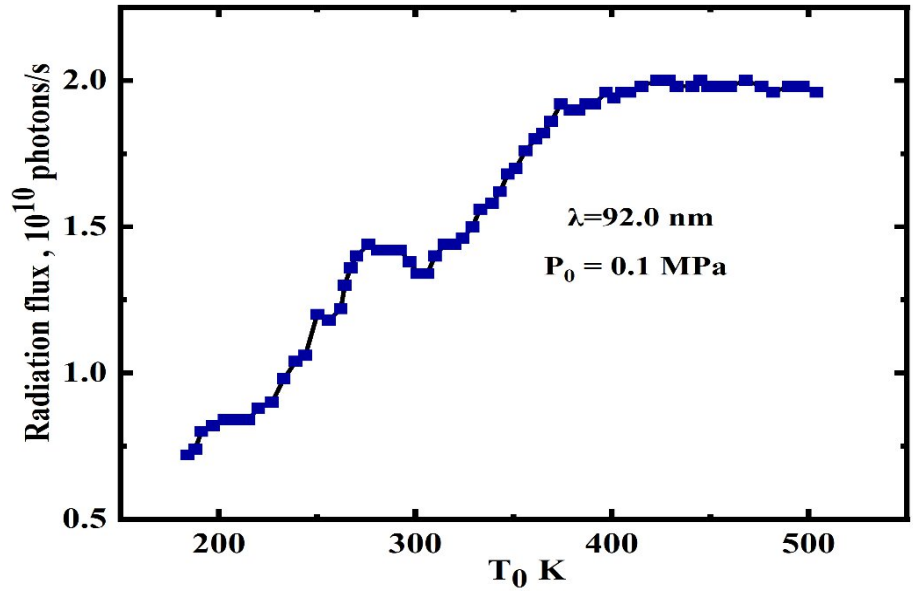

Fig. 2. Dependence of the Ar II line ($\lambda$=92.0 nm) radiation flux on the temperature and pressure of the gas at the nozzle inlet, $P_0 = 0.1$ MPa. Distance from the nozzle outlet section to the electron beam is 30 mm. Electron energy is 1 keV, and electron beam current is 20 mA.

It can be seen that in the atomic mode of the supersonic argon jet flow at $T_0 > 400$ K, the radiation flux of the $\lambda$= 92.0 nm line is $\Phi$ $(\lambda) = 2 \times 10^{10}$ photons/s. The dependence for line 93.2 is the same, with the radiation flux at $T_0 = 400$ K, $\Phi$ $(\lambda) = 1 \times 10^{10}$ photons/s. The relative error in determining the intensity of supersonic jet radiation in the studied wavelength range was approximately 20%.

It is known that the probability of electron emission during the excitation of a supersonic argon jet by electrons with fixed energy can be described using the photoemission cross section, determined by the formula (2):

$$\sigma(\lambda) = \frac{1}{n\, l\, \frac{i}{e}} \frac{4\pi}{\Omega} \Phi(\lambda), \#(2)$$

where $\sigma(\lambda)$ is the absolute photoemission cross section [cm$^2$], $n$ is the atomic jet density [cm$^{-3}$]; $l$ is the observed length of the electron beam in the excited jet zone [cm]; $\Omega$ is the solid angle of observation [sr]; $I$ is the electron beam current [A]; $e$ is the electron charge [C], $\Phi(\lambda)$ is the absolute radiation flux for the emission under study.

It should be noted that the contribution of cascade processes [14] to the intensity of the $\lambda = 92.0$ nm line does not exceed 1%, and self-absorption is insignificant. That allowed us to use the emission cross section $\sigma(\lambda)$ for the $\lambda = 92.0$ nm line at an exciting electron energy of 1 keV, measured earlier [14,15], and the absolute intensity $\Phi(\lambda = 92.0$ nm$) = 2 \cdot 10^{10}$ photons/s, to determine the concentration of argon atoms $n$ in the studied region of the jet at its atomic composition ($P_0 = 0.1$ MPa and $T_0 = 400$ K):

$$n = \frac{1}{\sigma(\lambda)\, l\, \frac{i}{e}} \frac{4\pi}{\Omega} \Phi(\lambda), \#(3)$$

As a result, at $P_0 = 0.1$ MPa and temperature $T_0 = 400$ K, the density of atoms in the studied area of the jet, determined by the diameter of the electron beam and the geometry of the jet relative to the entrance slit of the monochromator, is $n = 5.1 \times 10^{15}$ 1/cm$^3$.

The absolute intensities for the ArI resonance lines ($\lambda=104.8$ nm, $\lambda=106.7$ nm) measured at similar supersonic jet parameters $P_0 = 0.1$ MPa, $T_0 = 400$ K, and $L = 30$ mm have the following values:

- $\lambda=104.8$ нм $\Phi(\lambda) = 1.2 \times 10^{11}$ photons/s;
- $\lambda=106.7$ нм $\Phi(\lambda) = 0.36 \times 10^{11}$ photons/s.

Emission cross sections $\sigma(\lambda)$ for ArI resonance lines ($\lambda=104.8$ nm, $\lambda=106.7$ nm), using measured absolute intensities and at a density of atoms in the studied area of the jet n = 5.1 × 10$^{15}$ 1/cm$^3$, were determined by formula (2). The analysis showed that the obtained emission cross sections $\sigma(104.8$ nm$) = 1.7 \times 10^{-17}$ cm$^2$ and $\sigma(106.7$ nm$) = 0.52 \times 10^{-17}$ cm$^2$ significantly exceed the previously measured emission cross sections for argon resonance lines [14]. That can be explained by the fact that the measured intensities of the resonance lines $\lambda=104.8$ nm and λ=106.7 nm exceed the intensities corresponding to direct electronic excitation. Therefore, an assessment was made of the possible contributions of metastable states and cascade processes to the absolute intensity of these lines. Cascade contributions arise from higher excited states (e.g., $3p^54p$, $3p^55s$) decaying to 4s levels, filling the $^1P_1$ and $^3P_1$ states, which emit resonance lines λ=104.8 nm and λ=106.7 nm. The intensity from cascades depends on the excitation cross sections to these higher levels and their branching ratios to the 4s levels. A comparison of the emission cross sections obtained in this work with similar data [14] shows that the contribution of cascade processes to the measured intensities of resonance lines is approximately 50% for 104.8 nm and 40% for 106.7 nm. The metastable levels $3p^54s^3P_2$ and $^3P_0$ play an important role in discharge plasmas, contributing to the intensity of resonance lines through stepwise excitation. However, in a supersonic argon jet with a gas temperature of approximately 20 K and an atomic concentration of $5.1 \times 10^{15}$ cm$^{-3}$, these levels at an exciting electron energy of 1000 eV make a negligible contribution to the intensity of resonance lines compared to direct excitation by an electron beam. At an electron density in the electron beam of about $10^8$ 1/cm³, direct metastable pumping accounts for no more than 5%.

As can be seen in Fig. 2, at argon temperatures $T_0$ below 400 K, a decrease in the intensity of the 92 nm line caused by clustering is observed. At the same time, the spectrum shows the appearance of neutral $(Ar_2)^*$ λ=127 nm and charged excimer complexes $(Ar_4^+)^*$ λ=137 nm (Fig. 1) cluster continua, the intensity of which increases with decreasing temperatures. The dependence of the λ=127 nm continuum on temperature at P = 0.1 MPa in relative units is shown in Fig. 3.

The correlation analysis performed using the STATGRAPHICS plus software package revealed a strong inverse correlation ($r \approx -1$) between the temperature dependencies of the 92 nm ion line and the 127 nm cluster continuum (see Fig. 3, 4), which indicates a strong correlation between these dependencies: as the intensity of the cluster continuum increases with decreasing temperature (with increasing cluster size), a corresponding decrease in the intensity of the ion line is observed. The identified inverse correlation ($r \approx -1$) is a strong argument in favour of the non-cluster nature of the 92 nm line. Furthermore, in the cluster mode of supersonic argon flow, no broadening or shift of the 92 nm line relative to its position in the gas phase was observed, which may be caused by the influence of local fields or screening within the cluster. That is further evidence of the non-cluster nature of this line. Therefore, having determined the density of atoms in the jet in the absence of clustering, which on the curve for λ = 92 nm corresponds to a temperature of $T_0 = 400$ K, and establishing how this density changes with decreasing temperature, see Fig. 4, it was possible to calculate the fraction of condensate in the jet. That is defined as the ratio of the number of atoms that participate in cluster formation to the number of non-condensed atoms within the temperature range of 150–400 K. The calculated dependence of the condensate fraction on temperature is shown in Fig. 5.

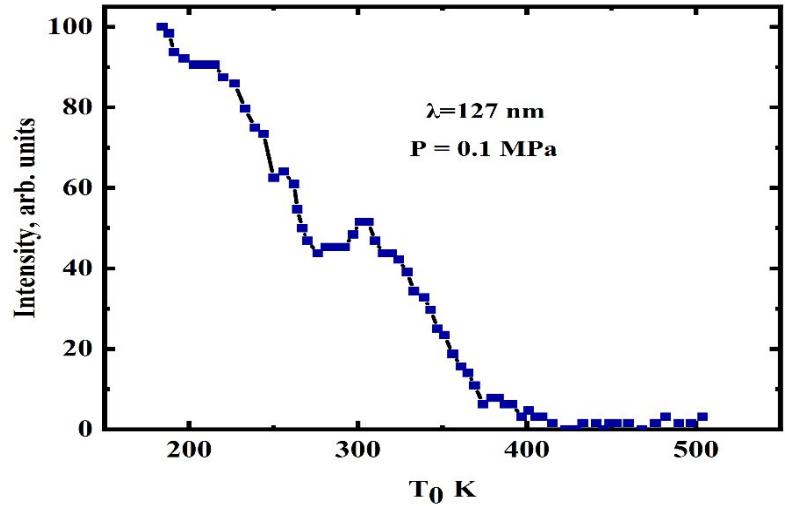

Fig. 3. Cluster continuum intensity λ=127 nm versus temperature $T_0$, at pressure $P_0 = 0.1$ MPa

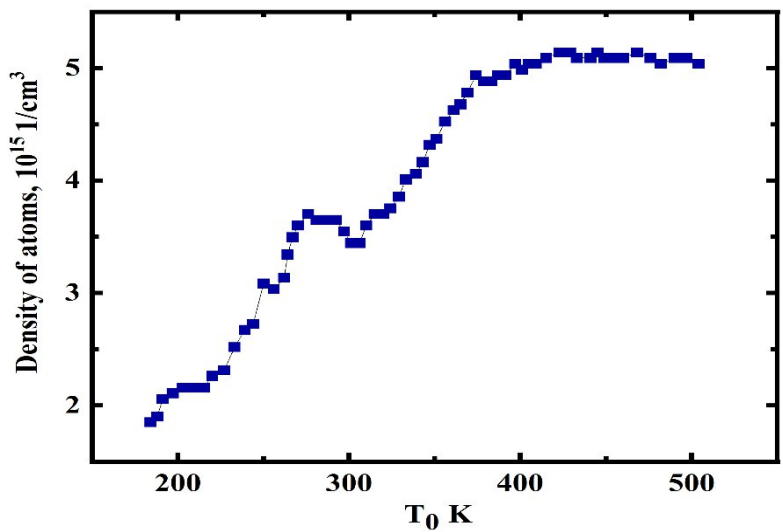

Fig. 4 Density of atoms in the investigated region of the supersonic jet

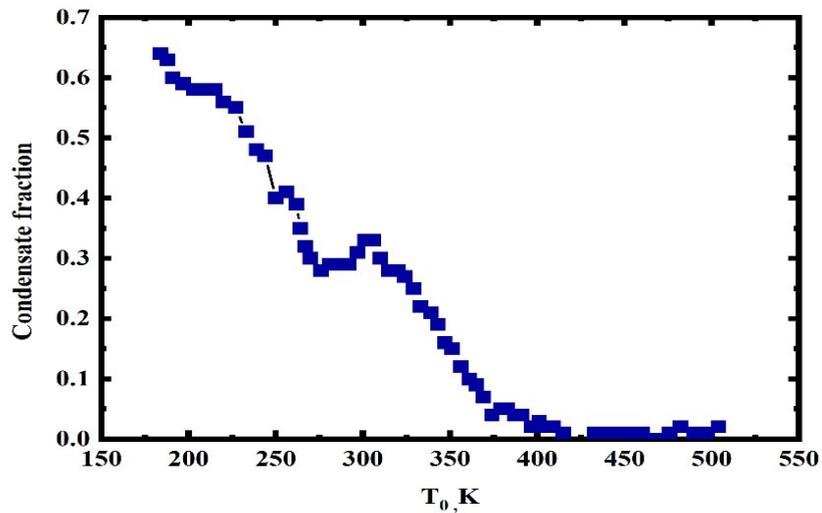

Fig. 5. The condensate fraction in a supersonic jet depending on the temperature of the gas at the nozzle inlet $T_0$ at a pressure of $P_0 = 0.1$ МПа.

Having estimated the number of atoms participating in clustering and the average cluster sizes experimentally determined at similar jet parameters [16,17], the density of clusters in the supersonic jet Ar as a function of their size was calculated (see Fig. 6).

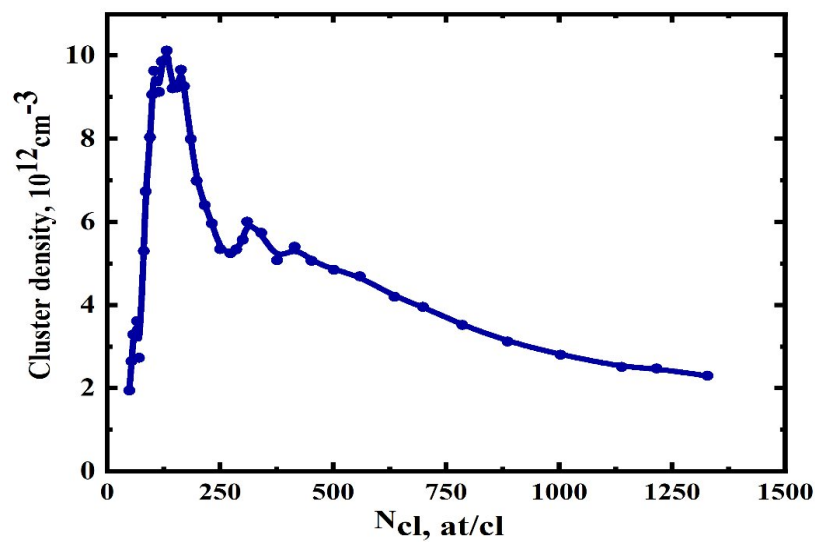

Fig. 6. The density of Ar clusters in the investigated region of the supersonic gas jet.

The curve shows a non-monotonic dependence of the density of clusters on their average size, which is characteristic of the systems in which cluster formation transitions from condensation processes dominated by nucleation to those dominated by coalescence. At the initial stage of the clustering process, the density of clusters increases rapidly as the number of condensation nuclei increases and reaches a maximum for an average cluster size of $N_{cl} \approx 150$ at/cl. Starting from $N_{cl} \approx 400$ at/cl, the coalescence process becomes predominant, increasing the average size of clusters and decreasing their number.

The obtained data on the density of clusters in a supersonic argon jet allow us to estimate the emission cross sections for the continuum $\lambda = 127$ nm emitted by clusters of different size and structure. However, since there are no experimental data on the size distribution of clusters at fixed parameters of the supersonic argon jet, the emission cross sections of the continuum can be estimated only for the average size of clusters. Thus, at $P_0 = 0.1$ MPa, $T_0 = 180$ K and an average cluster size of about 1000 at/cl, a relatively high emission cross section $\sigma$ ($\lambda$=127 nm) = $2.1 \cdot 10^{-16}$ cm$^2$ was obtained, indicating the effective conversion of the electron beam energy into the $\lambda$=127 nm continuum radiation.

### 3. Conclusion

This paper presents a new spectroscopic method based on the measurement of the integrated flux and spectral distribution of the radiation flux density of a supersonic argon jet in the range 50-150 nm. The method allows one to determine the density of the atomic component in the atomic and cluster modes of jet flow, the fraction of condensate, and the density of argon clusters at given parameters of jet flow in vacuum. The obtained parameters allowed us to determine emission cross sections for the argon ArI resonance lines ($\lambda$=104.8 nm and $\lambda$=106.7 nm) in the atomic mode of the jet outflow and the 127 nm continuum in the cluster mode at an average cluster size of 1000 at/cl. The emission cross sections can help understand the processes initiated by the electron beam in a supersonic argon jet. In the future, the use of the collision-radiation model developed for supersonic jets excited by an electron beam and including the energy levels of atoms, ions, and clusters will help to improve and make more accurate the methodology proposed in this work.